\RequirePackage{ifpdf}
\ifpdf 
\documentclass[pdftex]{sigma}
\else
\documentclass{sigma}
\fi

\newcommand{\mkmket}{|\frac{\ \, }{\ \, } k^- \rangle}
\newcommand{\pmkbra}{\langle ^+ \frac{\ \, }{\ \, } k|}

\newcommand{\PT}{{\cal P}{\cal T}}

\begin{document}

\allowdisplaybreaks

\renewcommand{\thefootnote}{$\star$}

\renewcommand{\PaperNumber}{043}

\FirstPageHeading

\ShortArticleName{The Analytic Continuation of the Lippmann--Schwinger
Eigenfunctions}

\ArticleName{The Analytic Continuation\\ of the Lippmann--Schwinger
Eigenfunctions,\\ and Antiunitary Symmetries\footnote{This paper is a contribution to the Proceedings of the VIIth Workshop ``Quantum Physics with Non-Hermitian Operators''
     (June 29 -- July 11, 2008, Benasque, Spain). The full collection
is available at
\href{http://www.emis.de/journals/SIGMA/PHHQP2008.html}{http://www.emis.de/journals/SIGMA/PHHQP2008.html}}}

\Author{Rafael DE LA MADRID}

\AuthorNameForHeading{R.~de la Madrid}

\Address{Department of Physics, The Ohio State University at Newark,
Newark, OH 43055 USA}

\Email{\href{mailto:rafa@mps.ohio-state.edu}{rafa@mps.ohio-state.edu}}

\ArticleDates{Received November 07, 2008, in f\/inal form March 30,
2009; Published online April 08, 2009}

\Abstract{We review the way to analytically continue the
Lippmann--Schwinger bras and kets into the complex plane. We will see that
a naive analytic continuation leads to nonsensical results in resonance
theory, and we will explain how the non-obvious but correct analytical
continuation is done. We will see that the physical basis for the
non-obvious but correct analytic continuation lies in the invariance of
the Hamiltonian under anti-unitary symmetries such as time reversal or $\PT$.}


\Classification{81S99; 81U15}

\renewcommand{\thefootnote}{\arabic{footnote}}
\setcounter{footnote}{0}

\section{Introduction}
\label{sec:Intro}

The Lippmann--Schwinger equation is one of the cornerstones of quantum
scattering. It was introduced by Lippmann and Schwinger in 1950~\cite{LS}, and
it has become standard in scattering theory~\cite{NEWTON,TAYLOR}.

For energies belonging to the scattering spectrum of the Hamiltonian,
the Lippmann--Schwin\-ger eigenfunctions describe scattering
states. In addition, one can describe resonances by analytically
continuing the Lippmann--Schwinger eigenfunctions into the resonance
energies. Such analytic continuation yields, up to a factor, the Gamow
states.

The purpose of this paper is to explain some of the subtleties of
the analytic continuation of the Lippmann--Schwinger eigenfunctions. We will
see that the naive\footnote{We will use the adjective ``naive'' for lack
of a better term.} analytic continuation is incompatible with
standard resonance
expansions, and we will therefore be forced to analytically continue the
Lippmann--Schwinger eigenfunctions in a somewhat counter-intuitive way. We
will f\/ind out that such counter-intuitive analytic continuation
is nevertheless physically sound, because it is rooted on the invariance
of the Hamiltonian under anti-unitary symmetries such as time reversal or
$\PT$. The hope is that this paper will clarify both the technicalities
of and the physical basis for the results of~\cite{06JPAII,08NPA}.

The structure of the paper is as follows. Section~\ref{sec:preliinaries}
is devoted to recall some well-known results of scattering theory. In
Section~\ref{sec:2pac}, we study both the naive and the correct analytic
continuations of the Lippmann--Schwinger
eigenfunctions. Section~\ref{sec:introduction} investigates the physical
soundness of such analytic continuations. In
Section~\ref{sec:conclusion}, we state our conclusions.

Throughout the paper, rather than working in a general setting, we shall use
the example of the spherical shell potential, although,
as explained in~\cite{08NPA}, the results hold for
a larger class of potentials that include, in particular, potentials of
f\/inite range.

\section{Preliminaries}
\label{sec:preliinaries}

Before proceeding with analytic continuations, it is convenient to brief\/ly
recall some basic results of scattering theory. For a more detailed account on
the results of this section, the reader is referred to~\cite{NEWTON,TAYLOR}.

\subsection[The Lippmann-Schwinger eigenfunctions]{The Lippmann--Schwinger eigenfunctions}

Let us take a simple example, such as the spherical shell potential:
\begin{gather}
           V(r)=\left\{ \begin{array}{ll}
                                0,   &0<r<a,  \\
                                V_0, &a<r<b,  \\
                                0 ,  &b<r<\infty   .
                  \end{array}
                 \right.
	\label{potential}
\end{gather}
For this potential, the Lippmann--Schwinger equation
\begin{gather}
       |E ^{\pm}\rangle = |E\rangle +
       \frac{1}{E-H_0\pm i\epsilon}V|E^{\pm}\rangle
       \label{LSeq}
\end{gather}
has the following solutions in the radial, position
representation for zero angular momentum:
\begin{gather}
      \langle r|E^{\pm}\rangle \equiv
       \chi ^{\pm}(r;E)= N(E)
     \frac{\chi (r;E)}{{\cal J}_{\pm}(E)}  ,
      \label{pmeigndu}
\end{gather}
where $N(E)$ is a $\delta$-normalization factor,
\begin{gather*}
   N(E) =
     \sqrt{\frac{1}{\pi}   \frac{2m/\hbar ^2}{\sqrt{2m/\hbar ^2   E } } }
  ,
\end{gather*}
$\chi (r;E)$ is the regular solution,
\begin{gather}
      \chi (r;E)=\left\{ \begin{array}{lll}
               \sin \Big(\sqrt{\frac{2m}{\hbar ^2}E }  r\Big), \quad &0<r<a,  \vspace{1mm}\\
               {\cal J}_1(E)e^{i \sqrt{\frac{2m}{\hbar ^2}(E-V_0) }  r}
                +{\cal J}_2(E)e^{-i\sqrt{\frac{2m}{\hbar ^2}(E-V_0) }  r},
                 \quad  &a<r<b, \vspace{1mm}\\
               {\cal J}_3(E) e^{i\sqrt{\frac{2m}{\hbar ^2}E }  r}
                +{\cal J}_4(E)e^{-i\sqrt{\frac{2m}{\hbar ^2}E }  r},
                 \quad  &b<r<\infty   ,
               \end{array}
                 \right.
             \label{LSchi}
\end{gather}
and ${\cal J}_{\pm}(E)$ are the Jost functions,
\begin{gather*}
     {\cal J}_+(E)=-2i{\cal J}_4(E)   ,
\qquad
      {\cal J}_-(E)=2i{\cal J}_3(E)   .
\end{gather*}
In terms of the Jost functions, the $S$~matrix is given by
\begin{gather}
      S(E)=\frac{{\cal J}_-(E)}{{\cal J}_+(E)}   .
     \label{SmatrixE}
\end{gather}
The functions ${\cal J}_{1}(E),\dots,{\cal J}_4(E)$ of equation~(\ref{LSchi}) are
determined by the conditions that $\chi (r;E)$ vani\-shes at $r=0$, and that
$\chi (r;E)$ is continuous with a continuous derivative at $r=a,b$. The
explicit expressions of ${\cal J}_1(E),\dots,{\cal J}_4(E)$ can be found in
\cite[Appendix~A]{JPA02}.

To the kets $|E^{\pm}\rangle$, there correspond the bras $\langle ^{\pm}E|$,
which are solutions of the following equation:
\begin{gather}
       \langle ^{\pm}E| =  \langle E| +
          \langle ^{\pm}E|V \frac{1}{E-H_0\mp i\epsilon}   .
       \label{LSeq-bras}
\end{gather}
Solving this equation in the radial, position representation yields the
``left'' Lippmann--Schwinger eigenfunctions:
\begin{gather}
       \langle ^{\pm}E|r\rangle = \overline{\chi ^{\pm}(r;E)}=
              \chi ^{\mp}(r;E)   .
      \label{leftLSeig}
\end{gather}

For the potential~(\ref{potential}), the Green function, the $S$ matrix and
the Lippmann--Schwinger eigenfunctions depend explicitly not on $E$ but on
the wave number $k$,
\begin{gather}
      k=\sqrt{\frac{2m}{\hbar ^2}E   } >0    .
      \label{wavenumber}
\end{gather}
This is why it is
convenient to work with $k$ rather than with $E$. In terms of $k$, the
Lippmann--Schwinger eigenfunctions (\ref{pmeigndu}) read as
\begin{gather*}
     \chi ^{\pm}(r;E)=
     \sqrt{\frac{1}{\pi}   \frac{2m/\hbar ^2}{k} }
     \frac{\chi (r;k)}{{\cal J}_{\pm}(k)}  .
   \label{kpmeigndu}
\end{gather*}
These eigenfunctions are normalized to $\delta (E-E')$. In order to normalize
them to $\delta (k-k')$, we def\/ine
\begin{gather*}
     \chi ^{\pm}(r;k) \equiv \sqrt{\frac{\hbar ^2}{2m} 2k } \chi ^{\pm}(r;E)
     =\sqrt{\frac{2}{\pi}}
     \frac{\chi (r;k)}{{\cal J}_{\pm}(k)}  .
\end{gather*}
It is important to note that in this equation $k$ is {\it positive}.

\subsection{The Riemann surface for the  square root function}

Analytic continuations are def\/ined for single-valued functions. When
the functions are not single-valued, the Riemann surfaces provide the means
to perform analytic continuations.

\begin{figure}[t]
\centerline{\includegraphics[width=14.3cm]{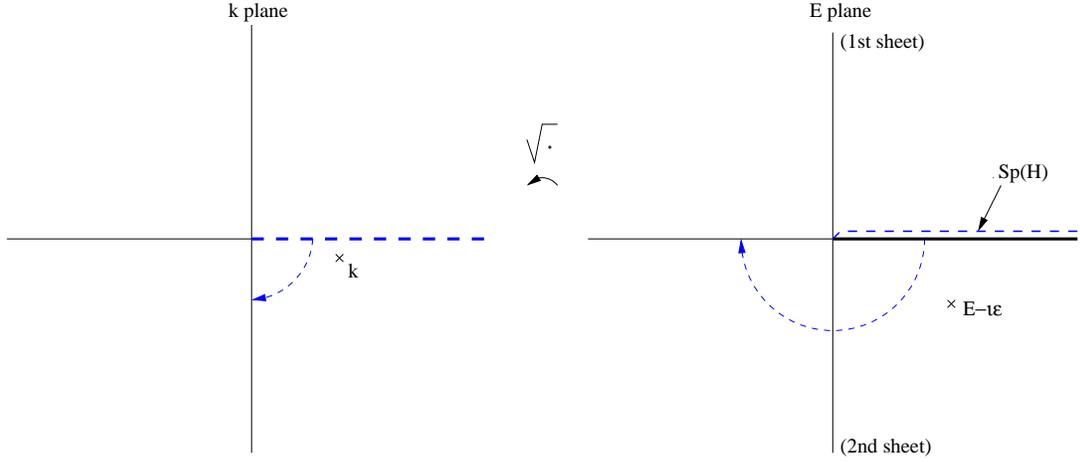}}
\caption{Analytic continuation from the upper rim of the cut into the lower
half plane of the second sheet: ``from above to below''.}
\label{minus}
\end{figure}

\begin{figure}[t]
\centerline{\includegraphics[width=14.3cm]{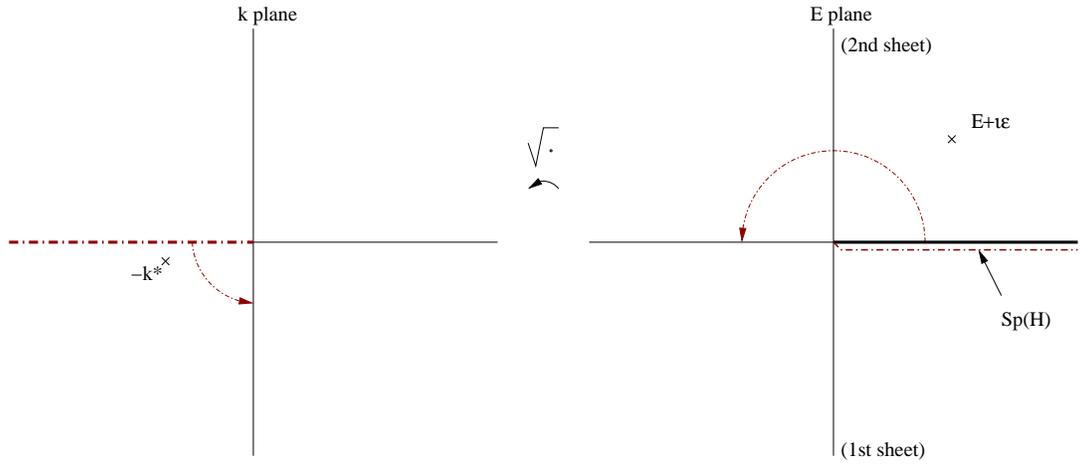}}
\caption{Analytic continuation from the lower rim of the cut into the upper
half plane of the second sheet: ``from below to above''.}
\label{plus}
\end{figure}

For the potential~(\ref{potential}), the multiple-valuedness of the Green
function, $S$ matrix and Lippmann--Schwinger eigenfunctions stems from the
square root function through equation~(\ref{wavenumber}). Thus, in our example,
in order to obtain analytic continuations, we need the Riemann surface
for the square root function. This Riemann surface is well known, and it has
been schematically displayed in Figs.~\ref{minus} and \ref{plus}. As shown
in these f\/igures, the
positive real axis of the energy plane corresponds to the spectrum
of the Hamiltonian. The upper rim of the positive $E$-axis maps onto
the positive $k$-axis, and the lower rim of the positive $E$-axis
maps onto the negative $k$-axis. The upper (respectively lower) half plane
of the f\/irst (respectively second) sheet maps onto the f\/irst
(respectively fourth) quadrant of the $k$-plane, see Fig.~\ref{minus}. The
upper (respectively lower) half plane of the second (respectively f\/irst)
sheet maps onto the third (respectively second) quadrant of the
$k$-plane, see Fig.~\ref{plus}.

The analytic continuation from the upper rim of the cut into the
lower half plane of the second sheet, that is, the continuation from the
positive $k$-axis into the fourth quadrant of the $k$-plane (see
Fig.~\ref{minus}), is sometimes called the ``from above to below'' analytic
continuation. Similarly, the analytic continuation from the
lower rim of the cut into the upper half plane of the second sheet, that is,
the continuation from the negative $k$-axis into the third quadrant of the
$k$-plane (see Fig.~\ref{plus}), is sometimes called the
``from below to above'' analytic continuation.

\subsection{The Gamow eigenfunctions}

The Gamow eigenfunctions are the state vectors of resonances~\cite{GAMOW,
SIEGERT,PEIERLS,HUMBLET,ZELDOVICH,BERGGREN,GASTON,BERGGREN78,SUDARSHAN,
MONDRAGON83,CURUTCHET,BL,LIND,BERGGREN96,BOLLINI,FERREIRA,CAVALCANTI,DIS,
BETAN,MICHEL1,AJP02,KAPUSCIK1,MONDRAGON03,MICHEL2,KAPUSCIK2,05CJP,MICHEL3,
MICHEL4,MICHEL5,URRIES,MICHEL6,TOMIO,COSTIN,ROSAS,MICHEL7}. They solve the
Schr\"odinger equation subject to purely outgoing boundary conditions. For
the spherical shell potential, the resonance energies come in complex
conjugate pairs $z_n$, $z_n^*$. The energy $z_n=E_n -i\Gamma _n /2$ is
the resonance pole, and it belongs to the lower half plane of the second
sheet of the Riemann surface. The energy $z_n^*=E_n +i\Gamma _n /2$ is the
anti-resonance pole, and it belongs to the upper half plane of the second
sheet of the Riemann surface. The corresponding resonance and anti-resonance
wave numbers are given by
\begin{gather*}
      k_n=\sqrt{\frac{2m}{\hbar ^2}z_n }  , \qquad
      -k_n^*=\sqrt{\frac{2m}{\hbar ^2}z_n^* }   , \qquad n=1,2, \ldots   ,
\end{gather*}
which belong, respectively, to the fourth and third quadrants of the $k$-plane.

In terms of the wave number $k_n$, the $n$th Gamow eigensolution
reads
\begin{gather*}
      u(r;z_n)=u(r;k_n)= N_n\left\{ \begin{array}{ll}
       \displaystyle  \frac{1}{{\mathcal J}_3(k_n)}\sin(k_{n}r),  &0<r<a, \vspace{1mm} \\
     \displaystyle    \frac{{\mathcal J}_1(k_n)}{{\mathcal J}_3(k_n)}e^{iQ_{n}r}
         +\frac{{\mathcal J}_2(k_n)}{{\mathcal J}_3(k_n)}e^{-iQ_{n}r},\quad &a<r<b,\vspace{1mm}
         \\
   \displaystyle      e^{ik_{n}r},  &b<r<\infty   ,
                           \end{array}
                  \right.
\end{gather*}
where $N_n$ is a normalization factor,
\begin{gather*}
       N_n^2=i \, \mbox{res} \left[ S(k) \right]_{k=k_n} ,
\end{gather*}
and where
\begin{gather*}
      Q_n=\sqrt{\frac{2m}{\hbar ^2}(z_n-V_0) }   .
\end{gather*}
The Gamow eigensolution associated with the $n$th anti-resonance
pole reads
\begin{gather*}
      u(r;z_n^*)=u(r;-k_n^*) =M_n\left\{ \begin{array}{ll}
       \displaystyle  \frac{1}{{\mathcal J}_3(-k_n^*)}\sin(-k_{n}^*r),  &0<r<a, \vspace{1mm}\\
       \displaystyle   \frac{{\mathcal J}_1(-k_n^*)}{{\mathcal J}_3(-k_n^*)}e^{-iQ_{n}^*r}
         +\frac{{\mathcal J}_2(-k_n^*)}{{\mathcal J}_3(-k_n^*)}
         e^{iQ_{n}^*r},\quad  &a<r<b ,\vspace{1mm}\\
         e^{-ik_{n}^*r},  &b<r<\infty   ,
                           \end{array}
                  \right.
\end{gather*}
where $M_n$ is a normalization factor,
\begin{gather*}
       M_n^2=i \, \mbox{res} \left[ S(k) \right]_{k=-k_n^*}=(N_n^2)^*    ,
\end{gather*}
and where
\begin{gather*}
       -Q_n^*=\sqrt{\frac{2m}{\hbar ^2}(z_n^*-V_0) }   .
\end{gather*}

\subsection{The analyticity structure of the ``total'' Green function}

\begin{figure}[t]
\centerline{\includegraphics[width=14.3cm]{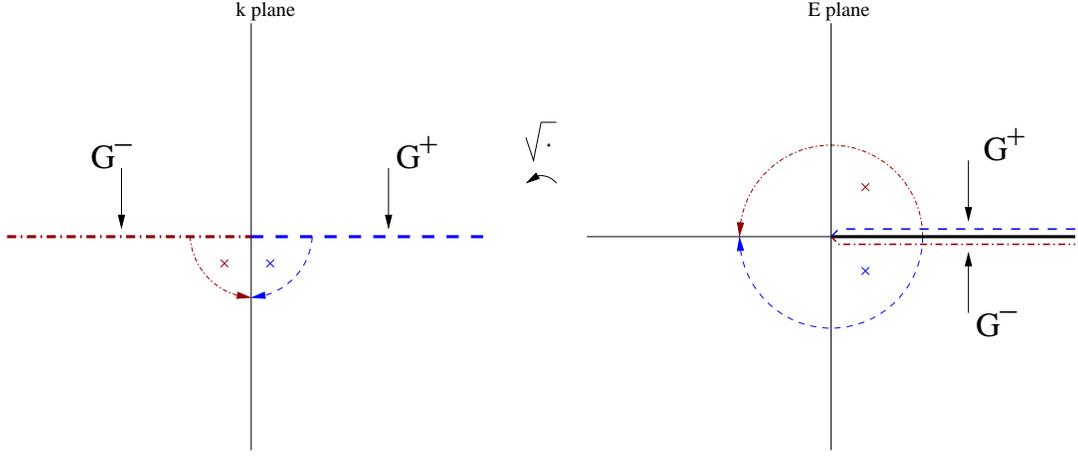}}
\caption{The ``from below to above'' and the ``from above to below''
analytic continuations of the Green function.}
\label{green}
\end{figure}

The Green function is the kernel of the resolvent when expressed
as an integral operator. In Dirac's bra-ket notation, it reads
\begin{gather*}
      G(r,s;z)=\langle r|\frac{1}{z-H}|s\rangle   ,
         \qquad z \in  {\mathbb C}   .
\end{gather*}
Since our potential does not bind bound states, the Green function is
analytic in the f\/irst sheet of the Riemann surface except for the positive
real axis, the cut, where it is discontinuous. Thus, when we approach the
cut from above we obtain a limit $G^+$ dif\/ferent from the limit $G^-$ that
we obtain when we approach the cut from below:
\begin{gather*}
      G^{\pm}(r,s;E) = \lim_{{\rm Im}(z) \to 0}G(r,s;z)   ,
       \qquad z=E\pm i  \, {\rm Im}(z), \qquad E,{\rm Im}(z) >0   .
\end{gather*}
The limits $G^{\pm}(r,s;E)$ are the well-known retarded and advanced
propagators, which in bra-ket notation are denoted as
\begin{gather*}
      G^{\pm}(r,s;E) =  G(r,s;E\pm i 0) =
 \langle r|\frac{1}{E-H \pm i \epsilon}|s\rangle   ,
          \qquad E\geq 0   ,
\end{gather*}
where $E\pm i 0 \equiv E\pm i \epsilon$ denote the energies on the upper
and lower rims of the cut. These propagators are related by complex
conjugation:
\begin{gather*}
     G^{+}(r,s;E) = \overline{G^{-}(r,s;E)}    .
\end{gather*}
Now, in order to obtain a resonance energy, we continue $G^{+}(r,s;E)$
``from above to below'' till we reach a pole in the lower half plane
of the second sheet, see Fig.~\ref{green}. Similarly, in order to obtain an
anti-resonance energy, we continue $G^{-}(r,s;E)$ ``from below to above''
till we reach a~pole in the upper half plane of the second sheet, see
Fig.~\ref{green}.

The above analytic continuations have counterparts in terms of the wave
number. If we denote the complex wave number by $q$, we
have that
\begin{gather*}
      G^{+}(r,s;k) = \lim_{{\rm Im}(q) \to 0}G(r,s;q)   ,
       \qquad q=k+ i \, {\rm Im}(q), \qquad  {\rm Im}(q),k >0   ,
\\
      G^{-}(r,s;-k) = \lim_{{\rm Im}(q) \to 0}G(r,s;q)   ,
       \qquad q=-k+ i \,  {\rm Im}(q), \qquad {\rm Im}(q),k >0   .
\end{gather*}
Thus, $G^+$ ($G^-$) is the boundary value of $G$ on the positive
(negative) $k$-axis. In terms of the wave number, the ``from above to below''
analytic continuation entails the continuation of $G^{+}(r,s;k)$ from the
positive $k$-axis into the fourth quadrant of the complex $k$-plane, and the
``from below to above'' analytic continuation entails the continuation of
$G^{-}(r,s;-k)$ from the negative $k$-axis into the third quadrant of the
complex $k$-plane, see Fig.~\ref{green}.

\subsection{The analyticity structure of the ``free'' Green function}

The ``free'' Green function,
\begin{gather*}
      G_0(r,s;z)= \langle r|\frac{1}{z-H_0}|s\rangle   , \qquad
         z \in {\mathbb C}   ,
\end{gather*}
has the same analyticity structure as the ``total'' Green
function, except that the analytic continuation of
the ``free'' Green function does not have any poles. The boundary values of
$G_0(r,s;z)$ on the upper and lower rims
of the cut are the so-called ``free'' retarded and advanced propagators,
\begin{gather}
      G_0^{\pm}(r,s;E)= G_0(r,s;E\pm i 0) =
       \langle r|\frac{1}{E-H_0\pm i \epsilon}|s\rangle   ,
        \qquad E\geq 0   .
      \label{rafreeGf}
\end{gather}

\subsection[The analyticity structure of the $S$ matrix]{The analyticity structure of the $\boldsymbol{S}$ matrix}
\label{sub:Smatrix}

The analyticity structure of the $S$ matrix~(\ref{SmatrixE}) is analogous to
those of the Green functions. As a function of the energy, $S(E)$ has a cut
through the positive real line. The boundary value on the upper rim of the
cut, $S(E+i0)$, is the complex conjugate of the boundary value on the lower
rim of the cut, $S(E-i0)$:
\begin{gather}
      S(E+i0) = \overline{S(E-i0)}   .
    \label{ccSuel}
\end{gather}
In the wave-number plane, equation~(\ref{ccSuel}) becomes
\begin{gather*}
      S(k) = \overline{S(-k)}   ,  \qquad k>0    .
\end{gather*}
The poles of the analytic continuation of $S(E+i0)$ ``from above to below'',
i.e., the poles of the analytic continuation of $S(k)$ into the fourth
quadrant, yield the resonance energies. The poles of the analytic
continuation of $S(E-i0)$ ``from below to above'', i.e., the poles of the
analytic continuation of $S(-k)$ into the third quadrant, yield the
anti-resonance energies.

The analytic properties of the $S$-matrix can be obtained through
very general arguments for a large class of potentials, see
e.g.~\cite[Chapter~2]{NUSSENZVEIG}. Essentially, one starts from the
values $S(E+i0)$ that the $S$
matrix takes on the upper rim of the cut, i.e., from the values $S(k)$, $k>0$,
that the $S$ matrix takes on the positive $k$-axis. One then continues
$S(E+i0)$ into both the f\/irst and second Riemann surfaces, i.e., one continues
$S(k)$ into the whole complex $k$-plane. Such continuation has the properties
mentioned above. As we shall see, the Lippmann--Schwinger bras and kets must
be continued this way, too.

\section[The boundary values and the analytic continuation of the
Lippmann-Schwinger bras and kets. Three resonance expansions]{The boundary values and the analytic continuation\\ of the
Lippmann--Schwinger bras and kets.\\ Three resonance expansions}
\label{sec:2pac}

After recalling the analyticity structure of the Green functions and the
$S$ matrix, we are in a~position to obtain the analyticity structure of the
Lippmann--Schwinger eigenfunctions.

\subsection{The naive, wrong approach}

\begin{figure}[t]
\centerline{\includegraphics[width=14.3cm]{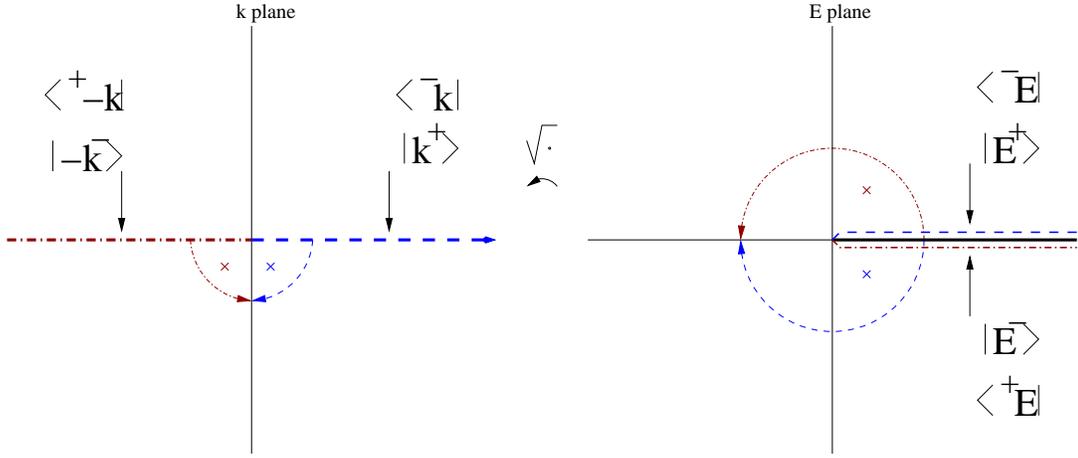}}
\caption{The naive analyticity structure of the
Lippmann--Schwinger bras and kets.}
\label{ls}
\end{figure}

Since the boundary values $G_0^{\pm}(r,s;E)$ enter the right-hand side of
the Lippmann--Schwinger equations~(\ref{LSeq}) and (\ref{LSeq-bras}), one
would naively expect that
the analyticity structure of the Green function carries over to the
Lippmann--Schwinger bras and kets. From the Lippmann--Schwinger
equations~(\ref{LSeq}) and (\ref{LSeq-bras}), from equation~(\ref{rafreeGf}), and
from the analyticity structure of the Green functions discussed in
Section~\ref{sec:preliinaries}, the ket $|E^+\rangle$ and the bra $\langle ^-E|$
seem like boundary values on the upper rim of the cut, whereas the ket
$|E^-\rangle$ and the bra $\langle ^+E|$ seem like boundary values on the
lower rim of the cut, see Fig.~\ref{ls}. These boundary values have
counterparts in the $k$-plane. The ket $|E^+\rangle$ and the bra
$\langle ^-E|$ correspond respectively to the ket $|k^+\rangle$ and the bra
$\langle ^-k|$, which seem like boundary values on the positive
$k$-axis, whereas the ket $|E^-\rangle$ and the bra $\langle ^+E|$ correspond
respectively to the ket $\mkmket$ and the bra $\pmkbra$, which
seem like boundary values on the negative $k$-axis, see Fig.~\ref{ls}.

From Fig.~\ref{ls}, it seems clear how one should analytically continue the
Lippmann--Schwinger bras and kets: The kets
$|E^+\rangle$, $|k^+\rangle$ and the bras $\langle ^-E|$, $\langle ^-k|$ should be
continued ``from above to below,'' whereas the kets $|E^-\rangle$, $\mkmket$
and the bras $\langle ^+E|$, $\pmkbra$ should be continued
``from below to above''.

However, as we are going to see through three resonance expansions, the
analyticity structure of the Lippmann--Schwinger bras and kets is not that of
Fig.~\ref{ls}.

\subsection{First resonance expansion}

The ``in'' bras and kets form a complete basis in the (formal) sense that
\begin{gather}
      1 =\int_0^{\infty}dE \, |E^+\rangle \langle ^+E|   ,
      \label{lsexpan}
\end{gather}
where $1$ is the identity operator. In terms of $k$, equation~(\ref{lsexpan})
reads as
\begin{gather*}
      1 =\int_0^{\infty}dk \, |k^+\rangle \langle ^+k|    .
\end{gather*}

\begin{figure}[t]
\centerline{\includegraphics[width=14.3cm]{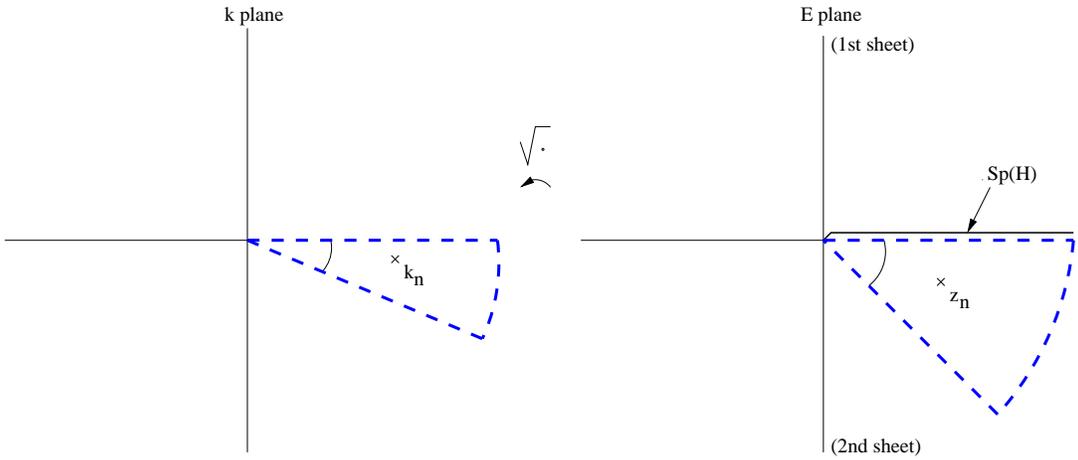}}
\caption{The contour to obtain resonance expansions. It is assumed that
the contour encloses all the resonances in the lower half plane of the
second sheet, and that the radius of the arc is sent to inf\/inity.}
\label{contour}
\end{figure}

The expansion~(\ref{lsexpan}) can be analytically continued into the complex
plane to obtain an expansion in terms of the Gamow
states. Using the contour of Fig.~\ref{contour}
to continue~(\ref{lsexpan}), we obtain an
expansion in terms of the Gamow states:
\begin{gather}
      1 = \sum_{n=1}^{\infty} |z_n^+\rangle \langle ^+z_n| + {\rm bg}   .
       \label{expaz}
\end{gather}
In terms of the wave number, equation~(\ref{expaz}) becomes
\begin{gather}
      1 = \sum_{n=1}^{\infty} |k_n^+\rangle \langle ^+k_n| + {\rm bg}   .
        \label{expak}
\end{gather}
The sum in equation~(\ref{expaz}) runs over all the resonance energies in the
lower half plane of the se\-cond sheet, and the sum in equation~(\ref{expak}) runs
over the corresponding resonance wave numbers in the fourth quadrant of the
$k$-plane. The symbol ``${\rm bg}$''
stands for the background integral. The ket~$|z_n^+\rangle$ (respectively $|k_n^+\rangle$) arises from the analytic
continuation of $|E^+\rangle$ (respectively~$|k^+\rangle$), and the bra
$\langle ^+z_n|$ (respectively~$\langle ^+k_n|$) arises from the analytic
continuation of $\langle ^+E|$ (respectively~$\langle ^+k|$).

If the naive analyticity structure of Fig.~\ref{ls} were correct, then in
using the contour of Fig.~\ref{contour} to obtain the above expansions,
the ket
$|E^+\rangle$ would be continued ``from above to below'' whereas the bra
$\langle ^+E|$ would be continued ``from {\it below} to {\it below}'',
i.e., from the lower rim of the cut into the lower half plane of the second
sheet, which makes no sense.

We note in passing that the expansion~(\ref{expaz}) must be obtained by
applying a regulator of the form $e^{-i\alpha E}$, $\alpha \to 0^+$, to both
sides of equation~(\ref{lsexpan}).\footnote{Physically, the regulator
parameter $\alpha$ plays the
role of time~\cite{05CJP}.} The reason is that the expansion~(\ref{expaz})
acquires meaning when we apply it to well-behaved wave functions. Since
in the energy representation such wave functions diverge exponentially,
we need the regulator $e^{-i\alpha E}$, $\alpha \to 0^+$, to kill such
exponential blowup. However, even if the wave functions didn't blow up
at inf\/inity, our conclusions would remain unaf\/fected, since we would be
nevertheless continuing $\langle ^+E|$ ``from {\it below} to
{\it below}''. Thus, in this paper we will not worry about such
exponential blowup at inf\/inity.

\subsection{Second resonance expansion}

The ``out'' bras and kets form another basis in terms
of which we can formally express the identity operator:
\begin{gather}
      1 =\int_0^{\infty}dE \, |E^-\rangle \langle ^-E|   .
      \label{lsexpan-}
\end{gather}
By way of the contour of Fig.~\ref{contour}, the analytic continuation of
(\ref{lsexpan-}) yields another resonance expansion:
\begin{gather*}
      1 = \sum_{n=1}^{\infty} |z_n^-\rangle \langle ^-z_n| + {\rm bg}   .
\end{gather*}

Now, if the naive analyticity structure of Fig.~\ref{ls} were correct, then
in using the
contour of Fig.~\ref{contour} to analytically continue~(\ref{lsexpan-}),
the ket $|E^-\rangle$ would be continued ``from {\it below} to
{\it below}'', which makes no sense.

We can also perform formal analytic continuations of the Lippmann--Schwinger
bras and kets into the upper half plane of the second sheet, thereby
obtaining the Gamow states associated with the
anti-resonances. In doing so, and if Fig.~\ref{ls} were correct, $\langle ^+E|$
and $|E^-\rangle$ would be continued ``from below to above'', whereas
$|E^+\rangle$ and $\langle ^-E|$ would be continued ``from {\it above} to
{\it above}'', i.e., from the upper rim of the cut into the upper half
plane of the second sheet, which does not make sense.

\subsection{Third resonance expansion}

From equations~(\ref{pmeigndu}) and (\ref{SmatrixE}), it follows that
\begin{gather}
      \langle r|E^+\rangle =S(E) \langle r |E^-\rangle   .
    \label{pkestmk}
\end{gather}
By substituting equation~(\ref{pkestmk}) into equation~(\ref{lsexpan}), we obtain
another completeness relation:
\begin{gather}
      1 =\int_0^{\infty}dE \, |E^-\rangle S(E) \langle ^+E|   .
         \label{expa2E}
\end{gather}
Using the contour of Fig.~\ref{contour}, the analytic continuation of the
expansion~(\ref{expa2E}) yields
\begin{gather}
      1 = \sum_{n=1}^{\infty} |z_n^-\rangle \langle ^+z_n| + {\rm bg}    .
           \label{expaz2}
\end{gather}
Thus, if the analyticity structure of Fig.~\ref{ls} were correct, then in
order to obtain~(\ref{expaz2}), $|E^-\rangle$ and
$\langle ^+E|$ would be continued ``from {\it below} to {\it below}''
rather than ``from above to below''. Such nonsensical ``from {\it below}
to {\it below}'' analytic continuation must be avoided.

\subsection[The Lippmann-Schwinger bras and kets as boundary values on the
upper rim of the cut]{The Lippmann--Schwinger bras and kets as boundary values\\ on the
upper rim of the cut}

\begin{figure}[t]
\centerline{\includegraphics[width=14.3cm]{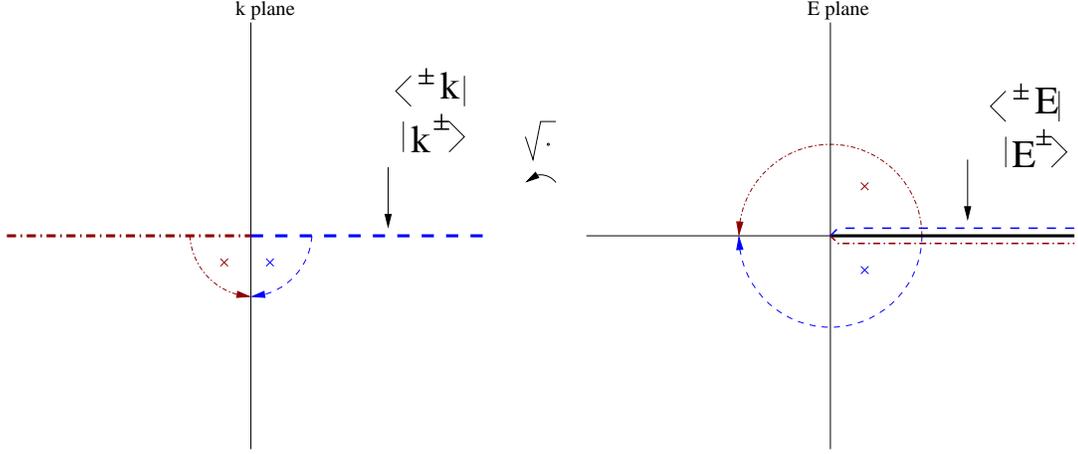}}
\caption{The Lippmann--Schwinger bras and kets as boundary values on the
upper rim of the cut, from which cut they are continued into the whole complex plane.}
\label{lsur}
\end{figure}

We have seen in the previous three subsections that if the Lippmann--Schwinger
eigenfunctions are to be boundary values of analytic functions, and if
Fig.~\ref{ls} depicts those boundary values correctly, then the
Lippmann--Schwinger eigenfunctions~(\ref{pmeigndu}) and (\ref{leftLSeig}) are
not always continued ``from above to below'' when we obtain resonance
expansions. It is clear that the above three resonance
expansions are obtained by analytically continuing the Lippmann--Schwinger
bras and kets ``from above to below'' only if the
Lippmann--Schwinger eigenfunctions~(\ref{pmeigndu}) and (\ref{leftLSeig})
are all boundary values on the {\it upper} rim of the cut, that is, they are
boundary values on the positive $k$-axis, see Fig.~\ref{lsur}.

In other words, if we denote the energies on the upper rim of the cut by
$E+i0$, then the Lippmann--Schwinger eigenfunctions take the following values
on that rim:
\begin{gather}
      \langle r|(E+i0)^{\pm}\rangle =
       \langle r|E^{\pm} \rangle = \chi ^{\pm}(r;E)   , \qquad E\geq 0   .
     \label{ularsmak}
\end{gather}
An analogous relation holds for the ``left'' Lippmann--Schwinger eigenfunctions:
\begin{gather}
     \langle ^{\pm}(E+i0)|r\rangle =
       \langle ^{\pm}E|r \rangle = \chi ^{\mp}(r;E)   , \qquad E\geq 0    .
    \label{ularsmab}
\end{gather}
Equations~(\ref{ularsmak}) and (\ref{ularsmab}) have natural counterparts
in the $k$-plane. If we denote the square root of $\frac{2m}{\hbar ^2}(E+i0)$
by $k$, $k$ being positive, then in the $k$-plane equation~(\ref{ularsmak}) reads as
\begin{gather}
      \langle r|k^{\pm}\rangle = \chi ^{\pm}(r;k)   ,  \qquad k\geq 0   ,
         \label{riefwnrepre}
\end{gather}
and equation~(\ref{ularsmab}) reads as
\begin{gather}
      \langle ^{\pm}k|r\rangle =  \chi ^{\mp}(r;k)    ,  \qquad k\geq 0   .
     \label{pmngkmos}
\end{gather}

The value that the Lippmann--Schwinger eigenfunctions take on {\it any}
complex energy is now obtained by analytic continuation of the
eigenfunctions~(\ref{ularsmak}) and (\ref{ularsmab}). In particular, the
continuation into the resonance energies is done by
continuing~(\ref{ularsmak})
and (\ref{ularsmab}) from the upper rim of the cut into the lower half-plane
of the second sheet. Likewise, the value that the
Lippmann--Schwinger eigenfunctions take on any complex
wave number is obtained by analytic continuation of the
eigenfunctions~(\ref{riefwnrepre}) and (\ref{pmngkmos}). In particular, the
continuation into the resonance wave numbers is done by
continuing~(\ref{riefwnrepre}) and (\ref{pmngkmos}) from the positive $k$-axis
into the fourth quadrant of the $k$-plane\footnote{A dif\/ferent
approach to analytically continue the Lippmann--Schwinger equation can
be found in~\cite{BG}.}.

Note that the analytic continuation into the anti-resonance poles, that
is, the ``from below to above'' analytic continuation, is already determined
by the values of the Lippmann--Schwinger eigenfunctions on the upper rim of the
cut. One needs f\/irst to analytically continue~(\ref{ularsmak}) and
(\ref{ularsmab}) from the upper to the lower rim of cut through the f\/irst
sheet of the Riemann surface. Once on the lower rim, one can continue into the
anti-resonance energies. In terms of the wave number, this means we are
continuing~(\ref{riefwnrepre}) and (\ref{pmngkmos}) from the positive to
the negative $k$-axis through the upper half plane. Once on the negative
$k$-axis, one can continue into the anti-resonance wave numbers.

It is worthwhile noting that the analytic continuation of the
Lippmann--Schwinger
eigenfunctions follows the same steps as the analytic continuation of
the $S$ matrix discussed in Section~\ref{sub:Smatrix}. One f\/irst
specif\/ies the boundary values on the upper rim of the cut (i.e., on the
positive $k$-axis), and then one obtains the values on any other complex point
by analytic continuation of those boundary values.

\section[Analytic continuation, time evolution, and anti-unitary symmetries]{Analytic continuation, time evolution,\\ and anti-unitary symmetries}
\label{sec:introduction}

In the previous section, we have seen that the analytic structure
implied by Fig.~\ref{ls} cannot be correct, because otherwise we would perform
nonsnensical analytic continuations of the Lippmann--Schwinger bras and
kets. In this section, we are going to see that, from a physical point of
view, the main reason why the analyticity structure
of the Lippmann--Schwinger eigenfunctions corresponds to the one depicted in
Fig.~\ref{lsur}, rather than to the one depicted in Fig.~\ref{ls}, is that the
time evolution of such eigenfunctions and their time reversal properties f\/it
into Fig.~\ref{lsur} rather than into Fig.~\ref{ls}.

In order to see why, let us consider a resonant system described by a
time-reversal inva\-riant Hamiltonian $H$. Since $H$ commutes with an
anti-linear operator, its complex eigenvalues must come in complex conjugate
pairs\footnote{As is well known, if a Hamiltonian $H$
commutes with an anti-linear operator, then $H$ has either real spectrum or
its complex eigenvalues come in complex conjugate pairs.}. One
complex energy is the resonance energy, whereas its complex
conjugate is the anti-resonance energy. Because the Gamow states satisfy
purely outgoing boundary conditions only for the resonance energies
$z_n$, whereas they satisfy purely {\it incoming} boundary conditions for the
anti-resonance energies $z_n^*$, the anti-resonances are interpreted as the
time-reversal of the resonances, see e.g.~\cite{SW,AJP02}. Since the
resonances
(respectively, anti-resonances) are obtained by analytic continuation from
the upper (respectively, lower) rim of the cut into the lower
(respectively, upper) half plane of the second sheet, forward time evolution
must be associated with the upper rim of the cut, whereas backward
(i.e., time-reversed) evolution must be associated with the lower
rim of the cut.

It is now clear why all the Lippmann--Schwinger eigenfunctions must be
def\/ined on the upper rim of the cut. Since the ``in'' eigenfunctions have
initial (i.e., prepared) boundary conditions built into them, whereas the
``out'' eigenfunctions have f\/inal (i.e., detected) boundary conditions
built into them, and since quantum scattering needs both initial and f\/inal
boundary conditions, both
the ``in'' and the ``out'' scattering states must appear in the time-forward
evolution of a scattering system, and therefore
both must be def\/ined on the upper rim of the cut. Their boundary values
on the lower rim of the cut are associated with backward time evolution. Thus,
the analyticity structure of the Lippmann--Schwinger eigenfunctions must be
given by Fig.~\ref{lsur}, not by Fig.~\ref{ls}.

We note that when the Hamiltonian $H$ is not invariant under time reversal
but instead is invariant under an anti-unitary symmetry such as
$\PT$~\cite{BENDER3,ALI1,ALI2,AHMED,ZNOJIL,ANDRIANOV,BENDER2,BENDER1}, the
above conclusions still hold. The only dif\/ference is that now the
anti-resonance poles are the $\PT$-reversed of the resonance poles, and that
the lower rim of the cut is associated with $\PT$-reversed time evolution.

\section{Conclusion}
\label{sec:conclusion}

A naive view of the Lippmann--Schwinger equation leads to the erroneous
conclusion that its solutions have the same analyticity structure as the
Green function. Such naive view leads to Fig.~\ref{ls} and to
the following table:

\vskip0.5cm

\centerline{\begin{tabular}{|l|l|l|l|l}
\cline{1-4}
{\bf Bra/Ket} & {\bf Expression} & {\bf Boundary value} &
{\bf Time evolution}  \\
\cline{1-4}
``in'' ket   & $\langle r|(E+i0)^+\rangle = \chi ^+(r;E)$ & upper rim &
forward   \\
\cline{1-4}
``out'' ket   & $\langle r|(E-i0)^-\rangle = \chi ^-(r;E)$ & lower rim &
backward  \\
\cline{1-4}
``in'' bra   & $\langle ^+(E-i0)|r\rangle = \chi ^-(r;E)$ & lower rim &
backward   \\
\cline{1-4}
``out'' bra   & $\langle ^-(E+i0)|r\rangle = \chi ^+(r;E)$ & upper rim &
forward  \\
\cline{1-4}
\end{tabular}
}

\vskip0.5cm

\noindent In the naive view, $|E^+\rangle$ and $\langle ^-E|$ are boundary
values
on the upper rim of the cut, and they are used solely for time-forward
evolution. For time-reversed evolution, one uses solely $|E^-\rangle$ and~$\langle ^+E|$, which are boundary values on the lower rim of the cut.

We have seen however that the Lippmann--Schwinger bras and kets should be
viewed as boundary values on the upper rim of the cut, from which rim they
can be analytically continued into the whole complex plane, much like the
continuation of the $S$-matrix is done from the upper rim of the cut into
the whole complex plane. All eigenfunctions are associated with
forward time evolution:

\vskip0.5cm

\centerline{\begin{tabular}{|l|l|l|l|l}
\cline{1-4}
{\bf Bra/Ket}  & {\bf Expression} & {\bf Boundary value} &
                                          {\bf Time evolution}  \\
\cline{1-4}
``in'' ket   & $\langle r|(E+i0)^+\rangle = \chi ^+(r;E)$ & upper rim &
forward   \\
\cline{1-4}
``out'' ket   & $\langle r|(E+i0)^-\rangle = \chi ^-(r;E)$ & upper rim &
forward  \\
\cline{1-4}
``in'' bra   & $\langle ^+(E+i0)|r\rangle = \chi ^-(r;E)$ & upper rim &
forward   \\
\cline{1-4}
``out'' bra   & $\langle ^-(E+i0)|r\rangle = \chi ^+(r;E)$ & upper rim &
forward  \\
\cline{1-4}
\end{tabular}}

\vskip0.5cm

\noindent Their analytic continuation into the lower rim of the cut is then used
in a time-reversed world:

\vskip0.5cm

\centerline{\begin{tabular}{|l|l|l|l|l}
\cline{1-4}
{\bf Bra/Ket}  & {\bf Expression} & {\bf Boundary value} &
                                          {\bf Time evolution}  \\
\cline{1-4}
``in'' ket   & $\langle r|(E-i0)^+\rangle = \chi ^+(r;E-i0)$ & lower rim &
backward   \\
\cline{1-4}
``out'' ket   & $\langle r|(E-i0)^-\rangle = \chi ^-(r;E-i0)$ & lower rim &
backward  \\
\cline{1-4}
``in'' bra   & $\langle ^+(E-i0)|r\rangle = \chi ^-(r;E-i0)$ & lower rim &
backward   \\
\cline{1-4}
``out'' bra   & $\langle ^-(E-i0)|r\rangle = \chi ^+(r;E-i0)$ & lower rim &
backward  \\
\cline{1-4}
\end{tabular}}

\subsection*{Acknowledgments}
\label{sec:ack}

The author wishes to thank the organizers of PHHQP~VII for their kind
invitation to participate in the conference and for their warm hospitality.

\pdfbookmark[1]{References}{ref}
\LastPageEnding

\end{document}